\documentclass[conference]{IEEEtran}
\usepackage[hyphens]{url}
\usepackage{hyperref}
\hypersetup{colorlinks,allcolors=blue}

\usepackage[backend=bibtex,style=ieee,citestyle=numeric-comp]{biblatex}

\usepackage{amsmath,amssymb,amsfonts}
\usepackage{graphicx}
\usepackage{textcomp}
\usepackage{xcolor}
\usepackage{booktabs}
\usepackage{array}
\usepackage{balance}

\title{Enhancing Attack Detection Capabilities in BACnet/IP Networks Using Machine-Learning Models}
\author{%
  \IEEEauthorblockN{Derek Manzella}
  \IEEEauthorblockA{%
    \textit{The Beacom College of Computer \& Cyber Sciences}\\
    \textit{Dakota State University}\\
    Madison, SD, USA\\
    derek.manzella@trojans.dsu.edu}
    \and
    \IEEEauthorblockN{John D. Hastings}
  \IEEEauthorblockA{%
    \textit{The Beacom College of Computer \& Cyber Sciences}\\
    \textit{Dakota State University}\\
    Madison, SD, USA\\
    john.hastings@dsu.edu}
}

\begin{document}

\maketitle
\begin{abstract}
Building Automation Systems (BAS) manage critical building functions using protocols such as BACnet/IP, yet defenders have limited tooling and few labeled datasets for detecting BACnet-specific attacks. This work addresses these gaps through three contributions. First, CISA's Zeek BACnet parser is modified to produce a unified per-packet log, simplifying feature engineering for machine-learning (ML) pipelines. Second, a simulated BACnet/IP testbed is developed using bacpypes3 to model a small commercial HVAC system with physics-based device behavior, schedule-aware controller logic, and per-packet attack labeling. Third, five unsupervised anomaly detection models are evaluated using baseline traffic and six BACnet attack types, including denial of service, reconnaissance, property tampering, and false data injection. Results show that One-Class SVM achieved the strongest overall performance, with an average F1 score of 0.864 across all attacks and F1 scores above 0.99 for high-volume denial-of-service and reconnaissance attacks. Detection is much stronger for high-volume attacks, such as DoS attacks and reconnaissance, than stealthier techniques such as tampering and false data injection, which scored around 77\%. 
\end{abstract} 

\begin{IEEEkeywords}
BACnet/IP, Building Automation Systems, Anomaly Detection, Machine Learning, Intrusion Detection Systems, Industrial Control System Security
\end{IEEEkeywords}

\section{Introduction}
Building Automation Systems (BAS) manage critical building applications such as HVAC, elevators, access control, fire and life safety, security systems, and mechanical systems. The Building Management System (BMS) market is expected to continue increasing each year \cite{mordor_bms_2026}. One of the most popular protocols used in BAS is BACnet \cite{GRAVETO2022102527, LI2023237}. A noteworthy trait among many BAS solutions and protocols is an inherent lack of security awareness. Many rely on isolation and a general lack of public awareness of protocol specifics \cite{GRAVETO2022102527, holmberg_2003_nistir7009}. Historically, many BAS have not been connected to the internet, which has helped protect them from these attacks. However, it is becoming more common for BAS networks to be used over the IP protocol and reachable via a corporate IT network. According to the search engine Shodan, there are roughly 31,000 internet-connected BACnet devices in the United States alone \cite{shodan}. The protocol's lack of focus on security has led to attackers having increased access to these networks via the internet and corporate IT networks \cite{9223152}.

There are a few different physical network layer options for BACnet, and this work focuses on BACnet/IP. BACnet is built to be highly interoperable, which increases an attacker's ability to communicate with BACnet devices once they gain access to a network. The original design of this protocol did not focus on security considerations, which allows for unauthorized access attacks, data interception, man-in-the-middle (MITM) attacks, Denial of Service (DoS), and malware. BACnet continues to add additional support to be more interoperable, such as supporting Internet of Things (IoT) devices. This further emphasizes the need to secure and detect network-based attacks against BACnet devices \cite{Narasimhan}.

There are two major classes of vulnerability in the BACnet protocol: lack of authentication and lack of encryption. Older standards and devices did not support any options for encrypted communications, which enables MITM attacks to intercept and modify data being sent between devices \cite{Narasimhan}. BACnet also lacks good authentication mechanisms, which allows for any device on the network to be able to read and write values of BACnet objects, as well as issue commands to devices. The protocol standard has added security-focused extensions in recent years, such as BACnet Secure Connect (BACnet/SC), which focuses on encryption. The problem, however, is that these features are rarely deployed in production environments \cite{kaur2015}. This results in a landscape where the majority of BACnet networks lack encryption and proper authentication, which makes network-level detection that much more important. 

Additional growth and usage of BACnet devices means that more and more traffic and data will be flowing through these networks. Standard logging, manual analysis, and static detections will become less effective at detecting sophisticated attackers with this increase in data and updates to protocol specifications and attack techniques. The amount of data to analyze manually does not scale well, and a better means of detecting anomalous and malicious activity in this protocol is necessary.    

This project proposes a unified framework for BACnet network data aggregation using Zeek, paired with real-time anomaly detection using various unsupervised machine learning (ML) models. The main goal of this research is to allow BACnet network defenders to better categorize standard baseline activity in BACnet networks while detecting anomalous activity, indicative of cyberattacks or mechanical failures. Static detections and rules are good at finding known and categorized attack types, but those known attack types are in short supply for the BACnet protocol. This is what ML excels at: generalizing the baseline activity in a network and automatically identifying anomalous activity. 

This work makes three contributions toward closing the security and detection gap that exists for the BACnet protocol. First, we modify the CISA icsnpp-bacnet Zeek protocol analyzer to combine all security-relevant network and protocol information into a single log, streamlining the feature engineering and data pre-processing steps for ML pipelines \cite{icsnpp_bacnet}. Second, we build a virtual BACnet/IP testbed using Docker and the bacpypes3 library that simulates a small commercial building’s HVAC system with schedule-aware traffic and dataset labeling functionality with the ability to simulate attack traffic. Third, we evaluate five unsupervised anomaly detection algorithms on the testbed datasets to identify which generalizes BACnet/IP activity best against six different attack types.

\section{Background}
\subsection{BACnet/IP Protocol Structure}
BACnet is the Building Automation and Control network protocol, designed by the American Society of Heating, Refrigeration and Air-Conditioning Engineers (ASHRAE). The protocol was first standardized back in 1995, and later became an ISO standard in 2004. The protocol is utilized in BAS in hundreds of thousands of buildings around the world \cite{ashrae135_2016}. 

\begin{figure}[ht]
    \centering
    \includegraphics[width=1\linewidth]{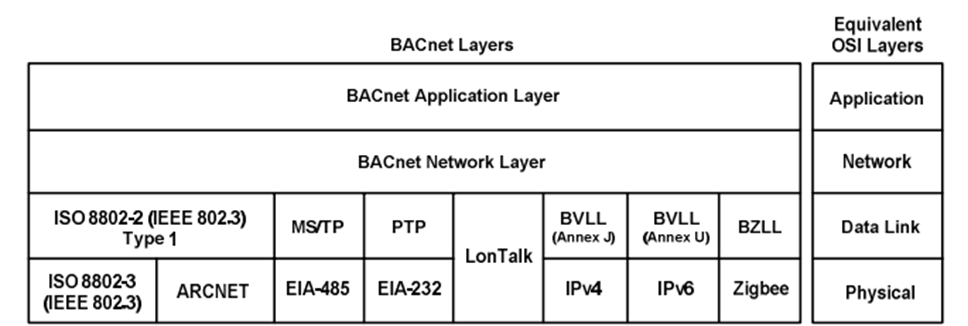}
    \caption{OSI model mapped to BACnet structure, highlighting the three core BACnet structures: BACnet Virtual Link Control (BVLC), Network Protocol Data Unit (NPDU), and Application Protocol Data Unit (APDU)}
    \label{fig:bacnet-struct}
\end{figure}

BACnet is based on a collapsed four-layer architecture, which corresponds to four of the OSI model layers. These four OSI layers are physical, data link, network, and application. Ethernet, ARCNET, MS/TP, PTP, LonTalk, BACnet/IP, BACnet/IPv6, and ZigBee are all supported options for the physical and data link layers used by BACnet \cite{ashrae135_2016}. Fig. \ref{fig:bacnet-struct} shows how the BACnet protocol structure corresponds to the OSI model.

BACnet packets consist of three main structures. The first is the BACnet Virtual Link Control (BVLC). This structure identifies the type of BACnet being used (0x81 for BACnet/IP), the function, and the length. The second structure is the Network-layer Protocol Data Unit (NPDU). This structure contains information related to the network communication of the connection, such as source and destination addresses, hop count, and message type. The third structure is the Application-layer Protocol Data Unit (APDU). This is where the security-relevant information for BACnet is located and contains the actual information being sent to and from BACnet devices.

\subsection{BACnet Objects, Services, and Properties}
BACnet devices use objects to represent the device's features and data types, which can be accessed across the network. The APDU is designed for network accessible devices to access and modify the properties and values stored on a specific device. The standard outlines 60 object types, including objects such as analog input and output, life safety zone, access zone, elevator group, and many more objects related to building controls \cite{10.1145/3140241.3140244, ashrae135_2016}. These objects are optional and are used based on each device's purpose. The device object is mandatory, however, and this contains information such as the vendor of the device and system status \cite{10.1145/3140241.3140244}.

Properties in BACnet reference the actual data being stored inside of every BACnet object. Each property consists of a data type, such as Boolean, integer, or an enumeration, and an access mode, which determines if it can be accessed or modified. For an HVAC network such as in our testbed, the present\_value property is used frequently to poll the current temperature reading of a sensor. Properties also carry metadata of objects, such as unit type and range of values. The standard also allows for vendor-specific objects to be used, extending the functionality of the protocol.

Services in BACnet are actions that can be used on a device. Each device implements a defined list of these services, which can be performed on the device. The standard outlines 46 BACnet services, with the main categories being alarm and event services, file access services, object access services, remote device management services, and virtual terminal services \cite{ashrae135_2016}. Services are typically used in a client-server architecture, where clients send service requests to servers, which reply to each request. Services allow client devices to read, write, and interact with objects linked to a device. 

\subsection{Threat Model}
There have been a few documented studies on attack types specific to Industrial Control Systems (ICS), as well as against the BACnet protocol. \textcite{moosavi2024developing} created a BACnet network traffic dataset using a simulated network of virtual machines (VMs) modeled on a real-world college campus. The study used three attack types in its dataset, the first being covert channel attacks. These highlight how the protocol can be used to send additional data along with valid packets, allowing for covert communications across BACnet traffic. The protocol also does not have any protections against DoS attacks, which enable an attacker on the network to degrade availability \cite{granzer_2010_security_analysis, moosavi2024developing}. The design of the protocol, which enables clients to issue commands and alter values of server devices, is another inherent vulnerability to the protocol, allowing for tampering of device values or injection of false data values to manipulate device behavior \cite{9223152, moosavi2024developing, 10.1145/3055186.3055189}.

There are additional security considerations when looking at cyberattacks against smart buildings. 
\begin{itemize}
    \item IoT device usage is becoming more common, increasing chances of internet exposure.
    \item Embedded devices have low resources, making complex encryption infeasible to implement.
    \item Cybersecurity is often an add-on feature, not baked in by default.
    \item Devices are vulnerable to physical attacks \cite{9223152}.
\end{itemize}

For this project, we assume an attacker already exists on the BACnet/IP network, either through lateral movement of a compromised IT environment or physical access to the BACnet infrastructure. The attacker device can send arbitrary BACnet messages to the network. The goal is to test and degrade availability, if possible, through DoS attacks, enumeration of the network and devices, as well as tampering and data injection to manipulate device functionality. This work focuses on network-based attacks; other categories of attacks, such as physical and supply-chain attacks, are not the focus of this study.

\section{Unified Data Pipeline}
\subsection{Limitations of ICSNPP-bacnet for ML}
Zeek is a popular open source tool for network data collection and processing. The Industrial Control Systems Network Protocol Parsers (ICSNPP) project from CISA \cite{kleinheider_2020_icsnpp} develops ICS protocol parsers for Zeek, focusing on network and security-relevant information specific to each protocol. The BACnet parser outputs four logs specific to BACnet traffic categories. A header log captures the header information of every packet analyzed by the sensor. The discovery log captures information about Who-is, Who-Has, I-Am, and I-Have message types. The property log captures information related to reading and writing of BACnet properties, which are the relevant fields to analyze and detect false data injection and tampering attacks. The device control log is focused on Reinitialize-Device and Device-Communication-Control message types. 

Having separate log sources is a hurdle to ML pipelines that process data from individual packets. In order to reconstruct the complete context of BACnet communications, the four separate logs need to be aligned and merged per-packet in order to successfully perform an ML pipeline on this data. For example, any BACnet request that modifies a property has connection and header information stored in bacnet.log, but the relevant application data, such as values, is stored in bacnet\_property.log. To streamline the process of data extraction, feature engineering, training, and testing of models, we created a modified parser to unify all log types into a single file and format.
\subsection{bacnet\_single Design}
Our bacnet\_single parser is a modified version of the icsnpp-bacnet parser from \textcite{icsnpp_bacnet} that consolidates all BACnet logs into a single file. None of the core parser functionality was changed; modifications only affected the logging functionality of the parser. We replaced the four record types with a single unified log that contains a union of all past fields. A new column in the log titled “log\_type” corresponds to the event categories from the original parser. A single logging stream outputs all BACnet events into bacnet.log. 

We verified the data parity between the original and new parser to ensure that all information was being accounted for and represented accurately in the new log output. We ran both parsers on two example PCAPs, and in both cases, each parser output the same number of total events. This ensures that all modifications made had no effect on the parsing functionality of the parser.

\section{BACnet Testbed}
\subsection{Architecture}

\begin{figure*}[ht]
    \centering
    \includegraphics[width=1\linewidth]{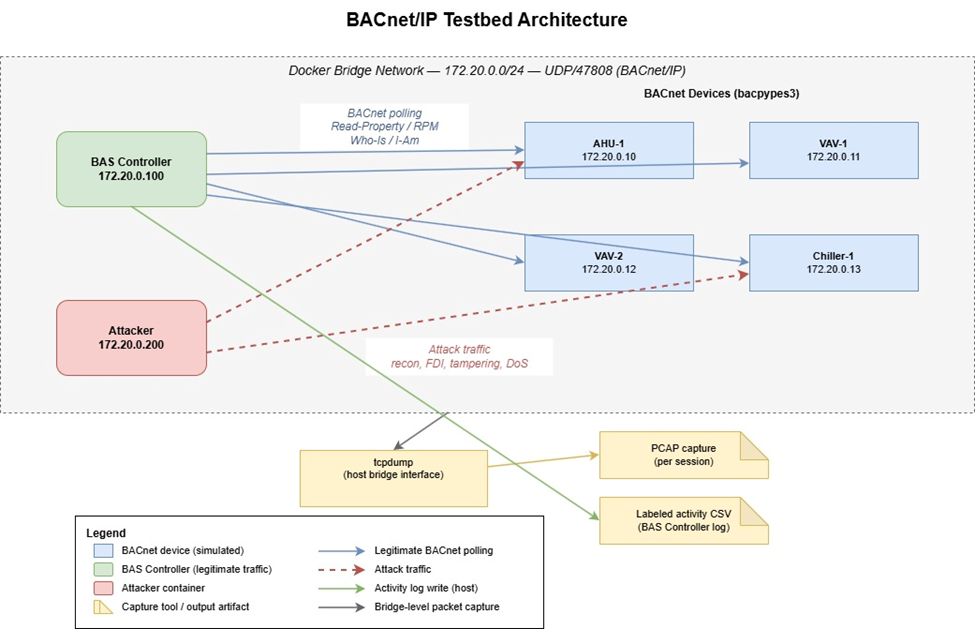}
    \caption{BACnet/IP testbed network map. Six containers were all located on a shared, private network of 172.20.0.0/24 on the default of UDP on port 47808. The controller manages the four BACnet HVAC devices, and the attacker system generates the attack traffic against the target devices.}
    \label{fig:testbed}
\end{figure*}

A Docker network was used to simulate a BACnet/IP network of a small-scale commercial HVAC system. The testbed runs on a Docker Compose stack with the six containers on a bridge network (172.20.0.0/24). All BACnet communication happened over the standard UDP port of 47808. Each device's role and addressing are listed in Fig. \ref{fig:testbed}. The BAS Controller polls and communicates with four separate field devices: a simulated Air-Handling Unit (AHU-1), two separate Variable-Air-Volume units (VAV-1, VAV2), and a Chiller unit (Chiller-1). The sixth container is the attacker system, which simulates a malicious actor who has network access to the BACnet devices. All malicious traffic in the datasets is from this device (172.20.0.200), which is later removed from training to prevent overfitting.

The Python bacpypes3 library \cite{bender_bacpypes3} was used to build the simulated device functionality expected from these BACnet devices. Each host has a unique IP address and sends and replies to BACnet traffic as if it were in a real network. The host system captures network traffic from the Docker bridge interface using the tcpdump utility. This generates the simulated BACnet PCAPs used and ingested by the models used for anomaly detection. 

\subsection{Traffic Simulation}

A priority for this testbed was to generate traffic that matched real-world BAS networks as closely as possible. The two main functions to create this were firstly a per-device simulated physics logic to control each device’s sensor values, and secondly, the BAS Controller communicating with each field device according to a simulated time of day schedule. 

The AHU computes a mixed-air temperature value based on simulated outside air, return air, and damper-position values stored as BACnet objects. Each VAV simulates a unique building zone controlling its respective temperature, which changes based on the supply air value it receives from the AHU. Sensor values of all devices fluctuate based on time of day (morning, evening, or overnight) to mimic real building behavior. When the supply temperature from the AHU increases or decreases, the downstream temperatures controlled by each VAV gradually shift to the set value, mimicking the real-world lag of physical devices.

No sensor values are static; each adds a minor amount of noise to prevent the baseline traffic from being too static and uniform. This adds a small amount of real-world variability to the devices and traffic to allow our models to better generalize to real-world networks.  

The BAS Controller is the device that manages all four HVAC systems by polling each device on scheduled intervals. In standard BACnet implementations, this is the device that controls the BACnet system devices by reading and writing their properties. The controller implements simulated time of day functionality, with four modes: Startup (6:30-7:00 AM), occupied (7:00 AM – 6:00 PM), shutdown (6:00 – 6:30 PM), and unoccupied (6:30 PM – 6:30 AM). Startup involves a high rate of Write-Property requests to initialize each of the four devices for the day, such as enabling the chiller device, elevating fan speeds, and setting value setpoints. Occupied mimics standard building operation polling of 30-second intervals to read sensor values and modify valve and fan values. Shutdown involves a high rate of Write-Property commands to slow down device functions by reducing fan speed values and modifying temperature setpoints. Unoccupied polls are at half the frequency of occupied (60-second intervals) and have much less broadcast messages, such as Who-Is requests.  

To speed up the generation of datasets, a time acceleration feature was implemented. Baselines were created with a time acceleration of 60, which translates to 1 second of real-time, generating 1 minute of BACnet network traffic. For example, a 30-minute traffic generation covers 30 simulated hours at this acceleration, which allows for a full time-of-day cycle through all four operating modes. An important note is that this acceleration affects the behavior of the network, but not the actual timings of packets being sent between devices. This allows for timings to be much more accurate and generalized to real networks, while allowing for easier generation of time-of-day network traffic patterns.

Together, these features enable the testbed to produce a variable dataset that never repeats the same traffic each run, and changes traffic characteristics through a simulated day. Packet rates, BACnet request frequency, and writes to objects vary depending on the current schedule mode with added noise. This allows for a more realistic environment and dataset for the models to train and learn on. It also allows for much greater flexibility in generating attack datasets for BACnet as well as simulating diverse attack types for ML training and evaluation. 

\subsection{Attack Scripts}

Attack scripts used in the testbed model some of the better-known and researched attack types against BACnet. The bacpypes3 library is used to interact with and attack the BACnet devices in this testbed. While traffic generation uses time acceleration to capture day and night traffic cycles, attacks are run without this to preserve realistic attack packet timings and responses. For example, each DoS flood attack and reconnaissance produces packets at a standard rate which matches real-world traffic. 

The six attack scripts used are described in the following table:
\begin{table}[htbp]
    \caption{Description of simulated testbed attacks.}
    \label{tab:attack_descriptions}
    \centering
    \renewcommand{\arraystretch}{1.4} 
    \begin{tabular}{|p{0.20\linewidth}|p{0.7\linewidth}|}
        \hline
        DoS Read & This attack floods active BACnet devices with Read-Property requests, sending around 100 packets per second\\
        \hline
        DoS Who-Is & This attack sends a high volume of Who-Is broadcasts, flooding the entire network with traffic and degrading availability. \\
        \hline
        FDI Replay & FDI attacks overwrite sensor values; the replay variant attack re-transmits previous baseline values from earlier. \\
        \hline
        FDI Sudden & FDI attacks overwrite sensor values; the sudden attack immediately changes a sensor value by a large amount. \\
        \hline
        Reconnaissance & Attacker issues many Who-Is broadcast messages to discover devices, followed by an enumeration of each BACnet device with Read-Property requests of each object on every responding device. \\
        \hline
        Tampering & Attacker sends targeted Write-Property requests to critical objects such as setting extreme temperature values, shutting off fans, and maxing out heat output. \\
        \hline
    \end{tabular}
\end{table}

\subsection{Labeling}
One of the limitations in the current state of research in BACnet security is a lack of labeled network traffic datasets. \textcite{moosavi2024developing} implemented simulated BACnet pcaps of baseline and attack traffic based on a real BAS network, but the dataset does not include per-packet labels, which makes it difficult to determine the accuracy and effectiveness of trained unsupervised ML models. \textcite{balamurugan_2023_cyber_faults_dataset} proposes a BACnet attack dataset, but this does not use network traffic. It uses application-level data generated by the Alfalfa ICS modeling tool \cite{alfalfa}, which makes it not an effective dataset for training a model on real-time BACnet network traffic.

These limitations led us to add labeling functionality to the testbed. Labeled datasets allow for accurate performance metrics from unsupervised models trained on the baseline dataset, as attack detection, false-positive rates, and true-positive rates can be determined. The testbed records a manifest of all traffic being generated for each session, saving the session ID, attack type, attack start time, and additional metadata. These labels are then appended to the bacnet.log file generated from the bacnet\_single parser, which can then be used for unsupervised training and supervised model evaluation.

\section{Methodology}
\subsection{Datasets}
There are two primary public BACnet attack datasets available. \textcite{moosavi2024developing} created a simulated dataset on a real BAS network, which contains 6 PCAPs. There is a baseline pcap, with five additional pcaps, each representing a unique attack type. These attacks include three pertaining to covert channel attacks abusing the protocol to hide data being sent across the network. The last two attacks are modifying and falsifying object values. These datasets are fairly large, with the biggest pcap being roughly 1.2 GB, which corresponds to over 45 million entries after being processed by the bacnet\_single Zeek parser. The size of the dataset, along with it being unlabeled, makes this dataset not an ideal candidate for unsupervised ML evaluation, since there is no way to verify which traffic in the dataset corresponds to the attack and therefore no way to accurately determine performance.

\textcite{balamurugan_2023_cyber_faults_dataset} from the National Renewable Energy Laboratory (NREL) creates another BACnet attack dataset, which is considerably smaller than that of Moosavi. The dataset is also not network traffic, which makes it incompatible with real-time network-based detections, which is the focus of this project. The Alfalfa ICS simulation tool \cite{alfalfa} was used to generate the dataset, which contains relevant application-layer information about BACnet device functionality. This dataset contains three baselines from three different times of year, as well as eight separate fault scenarios that focus on mechanical faults or failures rather than network attacks. Examples of these scenarios include:
\begin{itemize}
    \item Cooling Coil Valve Stuck Closed
    \item Heat Cool Operation without Min OA Damper
    \item Cooling Coil Valve Stuck Open
    \item OA Damper Stuck Open
    \item Low Supply Fan Speed
\end{itemize}

Having a large, unlabeled network traffic dataset (Moosavi) and a labeled, application-layer, non-network dataset (Balamurugan) leaves a gap in the security research space of BACnet. There exists no labeled BACnet network traffic attack dataset that allows for individual packet detection. Our testbed network (Fig.~\ref{fig:testbed}) is designed to help with this gap.

\subsection{Feature Engineering}

Each bacnet.log generated by the bacnet\_single Zeek parser was used as a training or testing sample. Anomaly detection is performed on a per-packet basis, with each packet related to an attack being labeled accordingly. Both the baseline dataset used to train each anomaly detection algorithm and all attack datasets share an identical engineered features list, with the attack datasets having an additional vector of labels for every packet indicating if it was normal or attack activity. 

Due to the nature of the testbed setup, IPs could allow models to easily detect every attack by learning that every packet from the attacker IP is anomalous. Due to this limitation of making a generalized model, IP addresses and ports were not used for these models. When using ML detection in a real-world network, this would be a great indicator to have, because having the algorithm baseline the activity from every IP on the network would greatly increase the performance of these models in detecting anomalous BACnet activity. We do not rely on this feature in our models here so we can determine detection based on BACnet protocol behavior.

\textbf{Temporal Features:}
 Temporal features were created using a 1-second window. The first feature was a packet count tracking total packets over a 1-second window. The second feature tracked the number of seconds from the last received packet. The third temporal feature tracked the total number of Write-Property or Write-PropertyMultiple requests made in the last second. The fourth feature tracks the value difference between consecutive packets with a number in the value field. These were grouped by a tuple of source host, object type, and instance number to ensure that the value difference referenced reading and writes of the same object value rather than from different devices or objects.

\textbf{Sparse and Presence Features:}
In normal BACnet communication, many of the available protocol fields and values are rarely, if ever, used. Many services and properties are never referenced, and fields such as password, result, vendor, and range are often never seen in standard communications. To reduce overall dimensionality, nine sparse fields were changed to a binary feature that checked only if the field was present in the packet. This removed any detection based on the content of these fields, but in the testbed baseline, they were infrequent enough for this trade-off to make sense. These nine features, which were changed to a binary feature, were result\_code, device\_id\_type, device\_id\_number, array\_index, time\_duration, password, result, vendor, and range. If training on a real network, it may be very beneficial to handle these differently if one of these fields is common in the environment. Two features changed to binary features are is\_orig which tracks the direction the packet was headed. Invoke\_id is also changed to a binary presence indicator, as well as instance\_number.  

\textbf{Categorical Features:}
Six fields were one-hot encoded. These fields were pdu\_service, pdu\_type, object\_type, property, log\_type, and bvlc\_function. Duplicate values of certain features that used different syntax of hyphens and underscores were merged to handle duplicate values. 

\textbf{Feature Removal:}
Eleven Zeek BACnet fields were dropped from the dataset prior to training and testing. These include the uid field, the four id\_* fields which track connection address information, the four fields for IP and port information,  and object\_name and device\_state. IP information had to be removed to prevent overfitting as the attacker's system address only generated malicious packets. 

\textbf{Summary:}
The final feature set consisted of 41 features, including a combination of temporal, binary, one-hot-encoded, and presence indicators. Features were then scaled before training and testing using a StandardScaler.

\subsection{ML Models}

Five unsupervised anomaly detection models were chosen based on past promising research on performance \cite{AGYEMANG2024e02386}. These five algorithms used for unsupervised anomaly detection of BACnet network traffic are Local Outlier Factor (LOF), Elliptic Envelope, Isolation Forest, One-Class Support Vector Machine (One-Class SVM), and One-Class SVM using Stochastic Gradient Descent. All five models are trained and fit only on baseline traffic to learn traffic patterns, then detect anomalous packets when testing against attack datasets.

Each of the five models was tuned using a grid search to find the hyperparameters that performed the best for each model. Models were trained on the baseline, and the grid evaluated model performance using the F1 score against the FDI sudden attack. This dataset was chosen as it is one of the stealthiest attacks, and was one of the hardest for the models to accurately detect. The optimal hyperparameters found for anomaly detection of BACnet network data are shown in Table \ref{tab:model_results}.

\begin{table*}[htbp]
    \centering
    \caption{Hyperparameter tuning results and validation performance across evaluated anomaly detection models. Models are ordered by descending Validation F1-Score, with the highest score highlighted in bold.}
    \label{tab:model_results}
    \begin{tabular}{@{}llc@{}}
        \toprule
        \textbf{Model} & \textbf{Best Hyperparameters} & \textbf{Validation F1-Score} \\
        \midrule
        One-Class SVM & \texttt{gamma='auto', nu=0.01} & \textbf{0.7733} \\
        Isolation Forest & \texttt{contamination=0.05, n\_estimators=100} & 0.7295 \\
        Local Outlier Factor & \texttt{contamination=0.05, n\_neighbors=50, novelty=True} & 0.6753 \\
        SGD One-Class SVM & \texttt{nu=0.05} & 0.6618 \\
        Elliptic Envelope & \texttt{contamination=0.1} & 0.1716 \\
        \bottomrule
    \end{tabular}
\end{table*}

\subsection{Evaluation}

Primary metrics for model evaluation used are F1 score, recall, precision, training time, and testing time. Model prediction accuracy is determined from the combination of F1 score, recall, and precision; model performance is determined from the combination of training and testing time. F1 score is used as testing data is skewed with more attack traffic, which inflates raw accuracy scores. F1 allows for a balance of precision and recall, and is a standard metric for ML intrusion detection system performance when false positives have a high cost \cite{Sommer}.

Models were all unsupervised and trained on unlabeled baseline testbed network data, then tested against each attack dataset. Performance of models on each dataset was recorded, and overall performance was determined by averaging the performance across all attack datasets. 

To assess which features are most important for each anomaly detected, we also evaluate the feature importance using sklearn.inspection.permutation\_importance on the two best performing models for each attack dataset. This helps evaluate which features are effective, and which may be less effective. 

\section{Results}

We trained each of the five unsupervised anomaly detection models on the same baseline testbed traffic and evaluated them against all six attack datasets. Results are shown in Table~\ref{tab:individual_dataset_scores}, and the overall averages are shown in Table~\ref{tab:averaged_model_leaderboard}. 

\begin{table*}[ht]
    \centering
    \caption{Performance metrics of evaluated models across individual attack datasets. The highest F1-Score for each dataset is highlighted in bold.}
    \label{tab:individual_dataset_scores}
    \begin{tabular}{@{}llcccc@{}}
        \toprule
        \textbf{Dataset} & \textbf{Model} & \textbf{F1-Score} & \textbf{Precision} & \textbf{Recall} & \textbf{Test Time (s)} \\ 
        \midrule
        
        DoS Read & Isolation Forest & 0.0071 & 0.7740 & 0.0036 & 0.056 \\
                 & One-Class SVM & 0.9929 & 0.9863 & 0.9996 & 0.448 \\
                 & SGD One-Class SVM & \textbf{0.9930} & 0.9867 & 0.9995 & 0.006 \\
                 & Local Outlier Factor & 0.9924 & 0.9852 & 0.9997 & 0.588 \\
                 & Elliptic Envelope & 0.9914 & 0.9865 & 0.9963 & 0.011 \\ 
        \addlinespace
        
        DoS Who-Is & Isolation Forest & 0.0028 & 0.5169 & 0.0014 & 0.045 \\
                   & One-Class SVM & \textbf{0.9916} & 0.9835 & 0.9998 & 0.381 \\
                   & SGD One-Class SVM & 0.8068 & 0.9771 & 0.6871 & 0.006 \\
                   & Local Outlier Factor & 0.9913 & 0.9828 & 0.9999 & 0.500 \\
                   & Elliptic Envelope & 0.9896 & 0.9839 & 0.9955 & 0.009 \\ 
        \addlinespace
        
        FDI Replay & Isolation Forest & \textbf{0.7268} & 0.7500 & 0.7051 & 0.009 \\
                   & One-Class SVM & 0.6841 & 0.5623 & 0.8733 & 0.025 \\
                   & SGD One-Class SVM & 0.5817 & 0.5151 & 0.6682 & 0.000 \\
                   & Local Outlier Factor & 0.6242 & 0.4650 & 0.9493 & 0.064 \\
                   & Elliptic Envelope & 0.2759 & 0.3664 & 0.2212 & 0.001 \\ 
        \addlinespace
        
        FDI Sudden & Isolation Forest & 0.7295 & 0.7740 & 0.6898 & 0.008 \\
                   & One-Class SVM & \textbf{0.7733} & 0.6513 & 0.9514 & 0.021 \\
                   & SGD One-Class SVM & 0.6618 & 0.6058 & 0.7292 & 0.000 \\
                   & Local Outlier Factor & 0.6753 & 0.5207 & 0.9606 & 0.078 \\
                   & Elliptic Envelope & 0.1716 & 0.2632 & 0.1273 & 0.002 \\ 
        \addlinespace
        
        Reconnaissance & Isolation Forest & 0.0132 & 0.2449 & 0.0068 & 0.012 \\
                       & One-Class SVM & \textbf{0.9589} & 0.9259 & 0.9943 & 0.056 \\
                       & SGD One-Class SVM & 0.8829 & 0.9205 & 0.8483 & 0.000 \\
                       & Local Outlier Factor & 0.9420 & 0.8926 & 0.9972 & 0.136 \\
                       & Elliptic Envelope & 0.4497 & 0.8504 & 0.3057 & 0.002 \\ 
        \addlinespace
        
        Tampering & Isolation Forest & \textbf{0.8831} & 0.8481 & 0.9211 & 0.008 \\
                  & One-Class SVM & 0.7858 & 0.6804 & 0.9298 & 0.022 \\
                  & SGD One-Class SVM & 0.7451 & 0.6772 & 0.8281 & 0.000 \\
                  & Local Outlier Factor & 0.7165 & 0.5748 & 0.9509 & 0.058 \\
                  & Elliptic Envelope & 0.2421 & 0.3906 & 0.1754 & 0.001 \\ 
        
        \bottomrule
    \end{tabular}
\end{table*}

\begin{table*}[ht]
    \centering
    \caption{Average performance metrics across all evaluated BACnet attack datasets. Models are ranked by descending average F1-Score, with the best overall score highlighted in bold.}
    \label{tab:averaged_model_leaderboard}
    \begin{tabular}{@{}lccccc@{}}
        \toprule
        \textbf{Model} & \textbf{F1-Score} & \textbf{Precision} & \textbf{Recall} & \textbf{Test Time (s)} & \textbf{Train Time (s)} \\ 
        \midrule
        One-Class SVM & \textbf{0.8644} & 0.7983 & 0.9580 & 0.1588 & 0.525 \\
        Local Outlier Factor & 0.8236 & 0.7368 & 0.9763 & 0.2373 & 0.500 \\
        SGD One-Class SVM & 0.7786 & 0.7804 & 0.7934 & 0.0020 & 0.016 \\
        Elliptic Envelope & 0.5200 & 0.6402 & 0.4702 & 0.0043 & 1.792 \\
        Isolation Forest & 0.3938 & 0.6513 & 0.3880 & 0.0230 & 0.238 \\
        \bottomrule
    \end{tabular}
\end{table*}

The strongest detector across all datasets is One-Class SVM, which has an average F1 score of 0.864 with an average precision of 0.798 and a recall of 0.958. Local Outlier Factor was a close second with an F1 score of 0.824, a precision of 0.737, and a recall of 0.976. SGD One-Class SVM performs fairly worse than OCSVM, but trains and performs much faster, which may be a tradeoff worth noting for high-throughput networks. The other models do not perform well on this dataset.

The main pattern from these results is the split between high-volume attacks compared to stealthier attacks (tampering, FDI). The top two models score above 0.97 in all three metrics for high-volume attacks, and have a harder time at predicting stealthier attacks accurately, with metric scores around the 0.8 range.

\section{Discussion}

The unified Zeek parser and labeled BACnet testbed data generation tool address two limitations in BACnet security research. The first is a general lack of publicly available BACnet network and attack datasets that work for real-time ML detections. Existing datasets are either unlabeled (Moosavi) or do not contain network traffic (Balamurugan), which makes it difficult to build and train accurate ML models. The second limitation is that the standard ICSNPP BACnet parser generates logs split between four logs, which adds an additional hurdle to a data pipeline for ML. The unified parser removes this limitation by seamlessly outputting all relevant information into a single, ready-to-go log file. 

Detection performance differs vastly based on the attack type. High-volume attacks (DoS and recon) are picked up easily by many of the anomaly detection algorithms, by using a one-second window to track packet counts over time. The best-performing models achieved F1, precision, and recall values near 0.99. This shows how powerful ML real-time detection capabilities can be for categorizing high-volume attacks. The anomaly detection models have a more difficult time categorizing stealthier attacks accurately. False data injection attacks were the hardest to detect by all models, with the best performance of the replay attack being an F1 score of 0.7268, with lower recall and precision scores across all models. These attacks send very little traffic, so further feature engineering is required in order to detect stealthy Write-Property requests that try to blend in with baseline traffic.

\section{Limitations}
The main limitation of this work is that the testbed environment is simulating a real-world BACnet/IP network, rather than using real-world data. While the testbed was designed with realistic scheduling, traffic patterns, and realistic sensors, simulated devices inherently cannot fully represent a production network with additional noise and variability, and this limited testbed does not represent a large and diverse BACnet network. This was a necessary trade-off due to the limited available datasets that can be used for this purpose. Many past papers in this field do not post their datasets, due to the data inside BACnet traffic containing sensitive information related to the building configurations. This work’s detection capabilities perform excellently at detecting high-traffic volume attacks, and perform worse at detecting stealthier and more targeted BACnet attacks. More research and fine-tuning are required in order to detect stealthier BACnet attacks at a high success rate. 

\section{Related Work}
\textbf{Threat Taxonomy:}
There have been a few studies on classifying and explaining various vulnerabilities and attack types against BACnet systems. These vulnerabilities include physical access attacks, internet exposure, modifying data values, injecting false data points, DoS, and firmware tampering \cite{10.1145/3140241.3140244, moosavi2024developing, granzer_2010_security_analysis, holmberg_2003_nistir7009}. This research motivated the threat model and attack types used for this paper.

\textbf{ML IDS for BACnet:}
\textcite{tonejc_2016_ml_bacnet, 7073181} both proposed ML detection specific to BACnet network traffic, using private data from internal BAS lab networks. The attacks they modeled included covert channel attacks, modifying APDU property values, DoS, and malformed packets. Both rely on private datasets, which are unavailable to the community, further motivating the labeled testbed from this project.

\section{Future Work}

This research focused on BACnet/IP implementations, but there are many other BACnet variants requiring visibility and detection capabilities, such as Master-Slave Token-Passing (MS/TP), and BACnet Secure Connect (BACnet/SC). Publicly available datasets using these other BACnet implementations are required in order for security systems to be built around available data. Additional attacks and baselines can be generated via our testbed in order to train and detect new attack techniques. This work, alongside other intrusion detection system research for BACnet traffic, can be integrated with security tooling such as Zeek to boost detection and visibility into BACnet networks using this popular and free open source tool. This capability can be added on top of the default logging of the tool, closing the gap between the security of IT and ICS systems. Further feature engineering can be done to better use network and application layer information to detect stealthier attacks, such as false data injection and tampering attacks. 

\section{Conclusion}

This project makes three contributions toward real-time ML-based intrusion detection systems for BACnet/IP network traffic. The first is a new, unified Zeek parser forked from the CISA/INL icsnpp-bacnet parser, which outputs all BACnet information into a single log, allowing for a single-pane view of all BACnet data and a streamlined data pipeline for ML. The second is a Dockerized testbed environment using bacpypes3 that simulates a small-scale HVAC building system with realistic sensors, time-of-day traffic patterns, added variability, and labeled attack dataset output for six attack types. The third is an empirical study of five unsupervised anomaly detection algorithms trained on baseline traffic and evaluated against six unique attack types.

Models performed very well at identifying and classifying high-volume attacks in the DoS read, DoS Who-Is, and reconnaissance attack datasets. Models struggled with correctly classifying stealthier false data injection and tampering attacks, but still were promising. The best performing anomaly detection model was found to be One-Class SVM, which averaged a total F1 score of 0.864 across all six attack types, with a precision of 0.798, and a recall of 0.958. Scores on the three high-volume attacks had an F1 score of around 0.99, with recall and precision scores also in the same range.

These models were intentionally built to try to generalize across all BACnet/IP implementations, rather than fit just our testbed network. Reintroducing IP address and port features based on each deployment should meaningfully improve overall model detection performance and accuracy.

\balance
\printbibliography
\end{document}